\documentclass[letterpaper,11pt]{article}
\usepackage{caption}
\pdfoutput=1 
\usepackage{xcolor}
\usepackage{enumerate}
\usepackage{framed}
\usepackage{mdframed}
\usepackage{amsmath}
\usepackage{amsfonts}
\usepackage{mathrsfs}
\usepackage{euscript}
\usepackage{amssymb}
\usepackage[normalem]{ulem}
\usepackage{graphicx, rotating}
\usepackage{epsfig}
\usepackage{latexsym}
\usepackage{graphicx}
\usepackage{color}
\usepackage{amsmath,bm,amssymb}
\usepackage{cite}
\usepackage{slashed}
\usepackage{hyperref}
\usepackage{epstopdf}
\usepackage[title]{appendix}
\usepackage[font=footnotesize,labelfont=]{caption}
\AppendGraphicsExtensions{.tif}
\hypersetup{colorlinks=true, citecolor=bluscuro, linkcolor=black, urlcolor=bluscuro}
\definecolor{rossos}{cmyk}{0,1,1,0.55}
\definecolor{bluscuro}{rgb}{0.15, 0.2, .85}
\definecolor{bluchiaro}{cmyk}{1,.3,0.,0.1}

\newcommand{\be}{\begin{equation}}
\newcommand{\ee}{\end{equation}}
\newcommand{\bea}{\begin{eqnarray}}
\newcommand{\eea}{\end{eqnarray}}
\newcommand{\beas}{\begin{eqnarray*}}
\newcommand{\eeas}{\end{eqnarray*}}

\newcommand{\rd}{{\rm d}}
\newcommand{\ri}{{i}}

\begin{document}
	\begin{center}
		
		~
		\vskip5mm
		
		{\Large  \bf{Modified Microcausality from Perturbation Theory
		}}
			\vskip10mm
			 Giordano Cintia$^a$,  Federico Piazza$^a$ and Samuel Ramos$^{a,b}$

		{\it  $^a$Aix Marseille Univ, Universit\'{e} de Toulon, CNRS, CPT, Marseille, France\\
		$^b$Universit\'e Paris-Saclay, CNRS, CEA,
Institut de Physique Th\'eorique, \\91191 Gif-sur-Yvette, France}
		
		\vskip5mm


	\end{center}
	
	\vspace{4mm}
	
	\begin{abstract}

Relativistic microcausality is the statement that local field operators commute outside the light-cone. This condition is known to break down in low-energy effective theories,  such as $P(X)$ models with a derivative interaction term of the ``wrong sign".  Despite their Lorentz-invariant form, these theories can exhibit superluminal propagation on  Lorentz-breaking backgrounds. We approach this phenomenon by computing the full operator-valued commutator in position space,   perturbatively in interaction picture.  After testing this formalism on a $\lambda \phi^4$ theory, we apply it to a $P(X)$ model. There, we show that the perturbative corrections to the free-theory commutator contain derivatives of delta functions with support on the standard Minkowski light cone.  While these corrections vanish on Lorentz-invariant states, they become ``activated" on states where Lorentz symmetry is spontaneously broken. In this case,  they approximate the new ``sound-cone" by means of a Taylor expansion. By applying linear response theory to an extended source, we show that deviations from standard causality are already present at first order in this expansion. Finally, we try to understand what goes wrong with the standard argument according to which Lorentz invariance implies microcausality.

 	 \end{abstract}

\newpage
\hrule 
\tableofcontents
 \vspace{0.6cm}
\hrule
\section{Introduction}
Commutators of local field operators are a direct probe of causality. In order to study their behavior, one typically looks at their \emph{expectation values} on some state or, equivalently, at the corresponding retarded and advanced propagators. These quantities are related by
\begin{equation}
G_c(t,\vec x) \, \equiv \, \langle [{\cal O}(t, \vec x), {\cal O}(0)]\rangle = G_R(t,\vec x) - G_A(t,\vec x)\, .
\end{equation}
In a Lorentz invariant setup and when ${\cal O}$ interpolates a light degree of freedom, $G_c$ has support on the light-cone,  $G_c (t,\vec x) \propto \delta (t^2 - {\vec x}^{\, 2})$. 
The situation is much more subtle in the absence of (manifest) Lorentz symmetry.  Of particular interest is the spontaneous breaking of Lorentz boosts on a state which  maintains approximate homogeneity and isotropy. This  pattern is ubiquitous in the real world, and applies equally well to cosmology and condensed matter physics~\cite{Nicolis:2015sra}. 
   In these situations the analytic properties of correlators have been the subject of interesting recent studies~\cite{Creminelli:2022onn,Heller:2022ejw,Hui:2023pxc,Creminelli:2023kze,Creminelli:2024lhd,Hui:2025aja}.

One possible consequence of spontaneously broken boosts is a speed of propagation different than the speed of light for some gapless  degrees of freedom. The simplest example is perhaps given by a  $P(X)$ theory,
\begin{equation} \label{ac_intro}
     \mathcal{L}=-\frac{1}{2}\partial_\mu \phi \partial^\mu \phi+\frac{\alpha}{4} \left(\partial_\mu \phi \partial^\mu \phi\right)^2\, .
    \end{equation}
When expanded around a time-dependent field configuration, $\phi(x) = \mu t + \varphi(x)$, the quadratic Lagrangian for the fluctuation $\varphi$ displays a speed of sound $c_s^2 \simeq 1 -  2\alpha \mu^2$. Therefore, by canonically quantizing $\varphi$, one obtains, to the lowest order in perturbation theory, a commutator for $\varphi$ that has support on the modified light-cone, \emph{i.e.} at $ \vec x^{\, 2} = c^2_s t^2 $~\cite{Hui:2025aja}.  

Notice that, as~\eqref{ac_intro} is Lorentz invariant, microcausality should apply to $\phi$-commutators as an operatorial statement. In particular, these commutators should vanish outside the standard Minkowski light-cone. The standard argument (see e.g.~\cite{Dubovsky:2007ac}) goes that, since boosts are implemented unitarily on the Hilbert space of the theory, commutators outside the lightcone are unitarily equivalent to commutators on equal time surfaces. If the latter vanish, also commutators at spacelike separation must vanish, which is the statement of microcausality. 
Then, how comes that, when evaluated on a Lorentz-breaking background, these objects become supported on a different light-cone? The question is particularly burning for $\alpha<0$, when the Lorentz breaking light-cone lies \emph{outside} the standard one. One might object that the choice $\alpha<0$  is unphysical, because the corresponding effective theory cannot be consistently completed in the UV~\cite{Adams:2006sv}.\footnote{In contrast, well-posed effective field theories obtained from Lorentz invariant UV-complete theories respect microcausality within their domain of applicability. See, for example, the fluid effective description of axion excitations in Weyl semimetals~\cite{Mottola:2023emy}.}  However, if we work exclusively at low energy and manage to leave these UV problems aside for the moment (see also~\cite{Kaplan:2024qtf} on this), we are still  facing a bizarre situation: the operator-valued commutator of $\phi$ must respect microcausality but seems to develop, on some states, expectation values that violate causality!

In this paper,  we compute the full operator-valued commutator in interacting theories,
perturbatively in interaction picture. We show that the emergence of a Lorentz-breaking light-cone is reconciled---at least formally---with Lorentz invariance by the derivatives of the delta function that appear in the perturbative calculation of the commutators in theories with derivative interactions. 
These delta derivatives are multiplied by combinations of field operators and are centered   on the Lorentz invariant light-cone at $\vec x^{\, 2} = t^2$ as expected from a Lorentz invariant theory. However, when evaluated on a Lorentz-breaking state, they manage to approximate the \emph{modified}  light-cone, in a  Taylor expansion  around the standard one, 
\begin{equation} \label{deltader_intro}
\delta(\vec x^{\, 2} - c_s^2 t^2) \ = \  \delta(\vec x^{\, 2} - t^2) + 2 \alpha \mu^2 t^2 \, \delta'(\vec x^{\, 2} - t^2)+ \frac{(2 \alpha \mu^2 t^2 )^2}{2}\,  \delta''(\vec x^{\, 2} - t^2) + \, \dots\, .
\end{equation}  
In the above, we have used that $1 - c_s^2 = 2 \alpha \mu^2$. 
To appreciate the effect, one does not need to resum the entire series. A modified propagation speed is already implied by the first, $\delta'$ correction, as we show in Sec.~\ref{sec_extended} by  applying linear response theory to an extended source. 

A similar mechanism must be at play in gravity, which is one motivation for the present study. By adopting the particle physicist's perspective advocated by Feynman and Weinberg among others, one can think of general relativity as the only consistent  Lorentz-invariant theory for a massless particle of helicity-2 at low energy. In this picture, any curved spacetime $g_{\mu \nu}$ is a (possibly, gigantic) perturbation of Minkowski space. The general covariant Einstein-Hilbert action $I_{EH} [g_{\mu \nu},\phi]$ in the presence of some field $\phi$ can thus be written as a Lorentz invariant theory of a scalar $\phi$ interacting with the helicity-2 field $h_{\mu \nu}$ in Minkowski,
\begin{equation} \label{EH}
I[h_{\mu \nu} , \phi] = I_{EH}[\eta_{\mu \nu} + h_{\mu \nu}, \phi]\, .
\end{equation}
In the above, $\eta_{\mu \nu}$ is the Minkowski metric. Once again, the $\phi$-commutators obtained from $I$ should satisfy Minkowski micro-causality. However, the causal structure that matters here is clearly the one inherited by the full metric $g_{\mu \nu}$. So the operator-valued commutators of $\phi$ in this theory should be ``flexible" enough to trace the general relativistic light-cone of some metric $g_{\mu \nu}$ once evaluated on the corresponding gravitational semiclassical state.\footnote{Something similar happens to transition amplitudes.  
The authors of the inspiring paper~\cite{Komissarov:2022gax} show that a cosmological FRW metric can be obtained from Minkowski space by adding a suitable background gravitational field. In this setup, the usual momentum-conservation of the S-matrix elements is reconciled with the cosmological gravitational redshift (or ``mode-stretching")  by a mechanism similar to the one displayed in~\eqref{deltader_intro} and based on derivatives of the Dirac-delta function.}  Whether or not such a light-cone is inside the Minkowski one is ultimately a gauge dependent statement, because it depends on how the points of Minkowski are identified with those of the full curved spacetime~\cite{Gao:2000ga}. While leaving gravity for future work, in this paper we only consider  scalar field theories and calculate the operator-valued commutator in a number of examples. 

This paper is structured as follows.  In Sec.~\ref{sec2}, we review the interaction picture formalism and show how it allows one to expand the operatorial commutator of an interacting scalar field theory in terms of the free field one. We validate the approach by applying it to the case of $\lambda \phi^4$ theory.  In Sec.~\ref{sec3}, we apply the method to a $P(X)$ theory, and show that the perturbative corrections to the free theory commutator involve derivatives of delta functions supported on the standard light cone. When computing expectation values, these contributions vanish on Lorentz-invariant states but persist when Lorentz symmetry is broken. By comparing with the semiclassical analysis, we show that these derivative terms reconstruct a modified sound-cone through a Taylor expansion. In Sec.~\ref{sec_extended}, we interpret these corrections as perturbative deformations of standard causality, using linear response theory with an extended source.  Sec.~\ref{sec5} concludes with a discussion of why the standard intuition that Lorentz-invariant theories yield commutators vanishing outside the light cone breaks down in $P(X)$ models with negative coupling.

\textit{Conventions and Notations.} We adopt the mostly-plus signature for the spacetime metric and set $\hbar=c=1$.  Three-dimensional spatial vectors are written as  $\vec{{x}}$ and $|\vec{{x}}|$ for their norm. In contrast, four-vectors read
 $x=(\vec{{ x}},t)$.
For quantum operators, we omit hats. Unless explicitly stated otherwise, all fields are assumed to be quantum fields. Heisenberg-picture fields are denoted by $\phi(x)$, while fields in the interaction picture are denoted by $\phi_0(x)$.  The four vector 0 indicates the origin of coordinate systems, so the pair $(\vec{{0}},0)$. If no spacetime label is provided, it is understood that all operators share the same coordinates.  Finally, we use the following convention for integral measures
\begin{equation}
\int_{t_1}^{t_2}d^4z=\int_{t_1}^{t_2}dz^0\int d^3z.
\end{equation}

\section{Commutators in the interaction picture formalism}
\label{sec2}
Microcasuality is the property that the commutator of fields is vanishing for space-like separated points
\begin{equation}
   [{\phi}(x),{\phi}(0)]=0, \qquad \text{if}\qquad x^2=t^2-\vec x^{\, 2} <0.
    \label{eq:microC}
\end{equation}

In interacting theories, the commutator is generally an operator-valued distribution, which can be analyzed perturbatively.  This makes microcausality an operatorial statement.  However, for a free massless scalar one simply gets 
\begin{equation}
  \Delta(x) \equiv [{\phi}_0(x),{\phi}_0(0)]= \text{sign}(t)\frac{1}{2 \ri \pi} \delta(t^2-\vec x^{\, 2})~,
\label{eq:commutatorFree}
\end{equation}
i.e., the commutator is a $c$-number with support on the light-cone.

In the following we show that operator-valued commutators in interacting theories can be evaluated in interaction picture, where they are expressed as convolutions of the interaction picture fields and of the free commutator~\eqref{eq:commutatorFree}.

\subsection{Interaction picture}
\label{Sect:interactionPict}
We start with a brief review of the interaction picture formalism \cite{Peskin:1995ev, Weinberg:2005vy}. We consider a Lagrangian density that can be split into a free and an interacting part, 
\begin{equation}
\mathcal{L}=\mathcal{L}_0+\mathcal{L}_{\rm int}\, .
\label{eq:generalTh}
\end{equation}
Although not necessary in general, in this paper the free Lagrangian is always that of a massless scalar field, 
\begin{equation}
\mathcal{L}_0 = -\frac12 \partial_\mu \phi \partial^\mu \phi.
\end{equation}
The Hamiltonian density of the theory is obtained in the standard way, 
\begin{equation} \mathcal{H}=\mathcal{H}_0[\phi,\pi]+\mathcal{H}_{\rm int} [\phi,\pi] \quad {\rm with} \quad \pi=\frac{\partial\mathcal{L}}{\partial\dot{\phi}}\, .
\end{equation}
Here, $\phi(x)$ and its conjugate momentum $\pi(x)$ are fields in the Heisenberg picture satisfying the standard equal-time commutation relations $[\phi(\vec{x},t), \pi(\vec{y},t)]=\ri \delta^{(3)}(\vec{x}-\vec{y})$. When considering derivative couplings,  the interaction Lagrangian $\mathcal{L}_{\rm int}$ may depend explicitly on the field velocity $\dot{\phi}$.  In such cases, the conjugate momentum $\pi$ becomes a nonlinear function of both $\phi$ and $\dot{\phi}$.  As a consequence, even the free Lagrangian $\mathcal{L}_0$, when expressed in terms of $\pi$, gives rise to interaction terms that contribute to the interacting Hamiltonian $\mathcal{H}_{\rm int}$. Therefore, it is crucial to define $\mathcal{H}_0$ as the quadratic part of the Legendre transform of $\mathcal{L}_0$.

The field in interaction picture is effectively a free field, as it evolves according to the free Hamiltonian $\mathcal{H}_0$. Hence, we call it $\phi_0(x)$.
 Introducing $t_*$ as the fiducial time at which the Heisenberg and interaction pictures coincide\footnote{In other words,
$$
   \phi_0(x)=e^{\ri H_0 (t-t_*)}\phi(\vec{x},t_*)e^{-\ri H_0 (t-t_*)}.
$$}, one finds that the Heisenberg and interaction fields are related by 
\begin{equation}
\phi(x)= 
U^\dagger(t,t_*)\phi_0(x)U(t,t_*),
\label{eq:HtoI}
\end{equation}
where $U$ is the \textit{interaction picture evolution operator}
\begin{equation}
    U (t_1,t_2) =T\,\text{exp}\left(- i\int_{t_2}^{t_1} d^4 z \,\mathcal{H}_I(z)\right)\, .
    \label{eq:timeordering}
\end{equation}
Here, $\mathcal{H}_I(z)$ stands for  $\mathcal{H}_{\rm int}[\phi_0(z), \pi_0(z)]$ and $T$ is the time ordering operator. 
The representation of the Heisenberg field of the interacting theory in terms of the interaction picture field, as provided by Eq.~\eqref{eq:HtoI}, is what allows us to evaluate the operatorial commutator. 

A set of similar relations applies to the conjugate momentum $\pi_0(x)$, which now satisfies the equal-time canonical commutation relations with $\phi_0(x)$.
In particular, we impose the following relation on $\pi_0$ at $t = t_*$,
\begin{equation}
\frac{\partial\mathcal{L}}{\partial\dot{\phi}}\Biggr\arrowvert_{t_*}=\pi{(\vec{x},t_*)}\equiv \pi_0(\vec{x},t_*)=\frac{\partial\mathcal{L}_0}{\partial\dot{\phi}}\Biggr\arrowvert_{t_*}\, .
\end{equation}
Notice that this implies that $\dot \phi \neq \dot \phi_0$ at $t_*$ if the interacting Hamiltonian is a function of the conjugate momentum.  As a result,  the total Hamiltonian cannot be consistently split into free and interacting parts if expressed in terms of the field and its time derivative. It is essential to rewrite it in terms of the field and its interacting conjugate momentum before performing the split.
See Ref.~\cite{Chen:2017ryl} for a concise review of the interaction picture formalism in the presence of derivative interactions.

By using the formalism depicted above, we evaluate the commutator 
$[{\phi}(x),{\phi}(0)]$, 
for a generic interacting theory of Lagrangian \eqref{eq:generalTh}. We start by expanding the evolution operators appearing in Eq.~\eqref{eq:HtoI} using the relation \cite{Weinberg:2005vy}
\begin{equation}
{\phi}(x)=\sum_{n=0}^\infty\,  { i}^n \int_{t_*}^t d^4 z_1 \, \dots \int_{t_*}^{t_{n-1}} d^4 z_n ~\big[\mathcal{H}_I(z_n),\big[\mathcal{H}_I(z_{n-1}), \dots [\mathcal{H}_I(z_1),\phi_0(x)\big]\big]\big].
\label{eq:FirstorderCommutator}
\end{equation}
By using the above formula, the Heisenberg picture commutator of the interacting fields can be obtained up to any desired order in the interaction Hamiltonian. In this work, we are mainly concerned with the first-order correction to the free commutator 
\begin{flalign}
[{\phi}(x),{\phi}(0)]=[\phi_0(x),\phi_0(0)]+{ i} \int_{t_*}^t {\rm}  d^4 z \big[\big[\mathcal{H}_I(z),\phi_0(x)\big],\phi_0(0)\big]-{ i} \int_{t_*}^0 d^4 z \big[\big[\mathcal{H}_I(z),\phi_0(0)\big],\phi_0(x)\big]
\label{eq:commutatorInt1}
\end{flalign}
The above  is the key equation we are using throughout the paper. 

The interaction picture formalism provides a solution for the interacting field in terms of the free one. An equivalent result can be obtained by solving the equation of motion for the field operator, convolving the interaction term with the Green's function of the quadratic theory. The field operator expansion given in Eq.~\eqref{eq:FirstorderCommutator} matches this latter method provided we choose the retarded Green's function, which is in constrast an implied choice in the interaction picture formalism. We demonstrate this equivalence using the Yang-Feldman formalism in Appendix~\ref{app_YF}.

\subsection{The example of $\lambda \phi^4$}

We apply the interaction picture formalism to study the commutator of a scalar field theory with a quartic local potential. 
Concretely, we introduce the following free and interacting Hamiltonian
\begin{equation}
    \mathcal{H}_0=\frac{1}{2}\pi^2+\frac{1}{2}\partial_i \phi\partial^i \phi,\qquad \mathcal{H}_{\rm int}=\frac{\lambda}{4!}\phi^4,
    \label{eq:quarticTh}
\end{equation} with $\pi=\dot{\phi}$. By plugging this interaction Hamiltonian in \eqref{eq:commutatorInt1}, we find 
\begin{equation}
[{\phi}(x),{\phi}(0)]=\frac{1}{2\pi \ri} \delta({\vec x}^{\, 2}-t^2)+\frac{\lambda}{8\pi^2 i} \int_{0}^t d^4 z \,\delta(z^2)\delta\left(({ z - x})^2 \right)  {\phi}_0(z)^2~.
\label{eq:firstorderQuartic}
\end{equation}
Here, we assumed $t>0$ and exploit the fact that the interaction picture commutator coincides with the commutator of the free theory \eqref{eq:commutatorFree}. By developing the above integral a little it is possible to see that the commutator has indeed no support outside the Minkowski light-cone.  By going to polar coordinates for $\vec z$, we start by integrating $\delta(z)$ in $d |\vec z\, |$,
\begin{flalign}
[{\phi}(x),{\phi}(0)]_{\mathcal{O}(\lambda)} =\frac{\lambda}{8\pi  \ri}\int_{0}^t z^0 { d}z^0 \int \frac{d\Omega}{2\pi} \delta\left(t^2-2 t z^0-{\vec x}^{\, 2} +2 |\vec x|z^0 \cos \vartheta\right) {\phi}^2_0(z)\bigg|_{{|\vec z\, |}=z^0}~.
\label{eq:quarti3}
\end{flalign}
Here, we focus only on the $\lambda$ correction to the free commutator. Moreover, we applied the relation
\begin{equation}
    \delta(z^2)=\frac{1}{2|\vec z\, |}\biggr\{\delta(|\vec z\, |-z^0)+\delta(|\vec z\, |+z^0)\biggr\}
    \label{eq:deltarelatin}
\end{equation}
and use the fact that the second delta function on the right-hand side has no support in the $z^0$ integration range considered.
Finally, we check what are the conditions under which the remaining delta function in~\eqref{eq:quarti3} has support. In particular, we have to check for what values of $|\vec x|$, $t$ and $z^0$, the cosine satisfies the relation $-1\leq\cos\vartheta\leq 1$.
One sees that this condition implies~\footnote{After some algebraic manipulations, deriving the relation below requires determining the conditions on $x$, $t$ and $z_0$ that satisfies \begin{equation}
    x^2-{t}^{\, 2} \lessgtr  2z^0(\pm|\vec{x}|  - t),
\end{equation}
where we have to restrict to $|\vec{x}|>0$, $t>0$ and $0<z^0<t$. With these restrictions, the lower inequality is satisfied if $x>t-2z^0$. In contrast, the upper inequality is instead satisfied if $x>-t+2z^0$. By overlapping these results with the restrictions, Eq.~\eqref{eq:DeltaSupport} follows. 
} 
\begin{equation}     \label{eq:DeltaSupport}
 |-t+2z^0|\leq |\vec{x}| \leq t\, .
\end{equation}
The restriction of the integral to $0\leq z^0 \leq t$, plays a key role in deriving \eqref{eq:DeltaSupport}.  If instead we were allowed to consider also $ z^0 \geq t $,  then the delta function could also have support if $\vec x^{\, 2}>t^2$. 

It is straightforward,  but tedious, to show that the vanishing of the commutator at space-like separation holds also at order $\lambda^2$. This is obtained by expanding the commutator to the second order in the interaction Hamiltonian, and check for what values of $|\vec{x}|$ and $t$ the $\lambda^2$ corrections has support.
\subsection{Evaluating the commutator on a homogeneous field configuration}

While a free massless field propagates strictly on the light-cone, adding self-interactions can extend the support of the commutator well \emph{inside} the light-cone, i.e. at time-like separation. This is somewhat implicit in~\eqref{eq:quarti3}. In order to make it more explicit we now evaluate this expression on a  coherent state describing a spatially homogeneous field configuration,
\begin{equation}
|{\bar \phi}\rangle=\text{exp}\left(-\ri {\bar \phi}\int d^3 x  \, \pi(\vec{x },0) \right)|0\rangle.
\end{equation}
Here, $|0\rangle$ is the Fock vacuum of the free theory, and ${\bar \phi}$ represents the expectation value of the field operator at $t=0$. 

Some properties of this state are discussed in App.~\ref{sec:COH}, and we refer the reader also to Refs\cite{Berezhiani:2020pbv,Berezhiani:2021gph}. 
 One property worth mentioning here is that the expectation value of the quadratic field operator on the coherent state reads
\begin{equation}
\langle \bar{\phi}|\phi^2_0(z)|\bar{\phi}\rangle=\bar{\phi}^{\,2}+\langle 0|\phi^2_0(z)|0\rangle
\label{eq:quadratic_correlation}
\end{equation}
to leading order $\lambda$.
By using this expression,   we thus need to evaluate
\begin{equation}
\langle {\bar \phi}|[{\phi}(x),{\phi}(0)]|{\bar \phi}\rangle=\frac{1}{2\pi \ri} \delta({\vec x }^{\, 2}-t^2)+\frac{\lambda {\bar \phi}^{\, 2}}{16 \ri \pi\, |\vec x| }\int_{0}^t  { d} z^0 \int_{-1}^1d \cos\vartheta ~\delta\left(\cos \vartheta-\frac{\vec x ^{\, 2}-t^2+2 t z^0}{2|\vec x|  z^0}\right) \, .
\label{eq:firstorderQuartic}
\end{equation}
Higher order corrections to~\eqref{eq:quadratic_correlation} give only ${\cal O}(\lambda^2)$ contributions to the above.  Moreover,  we removed the constant divergent term $\langle 0|\phi_0^2|0\rangle$ by renormalizing the theory.  This can be eliminated by introducing the standard vacuum mass counterterm in the interaction Hamiltonian~\eqref{eq:quarticTh}, that precisely cancels it out.

An interesting feature of this method is that when \textit{contanct potentials} are involved we can evaluate everything in position space by integrating over the $\delta$-functions. In other words, we do not need to refer to any momentum space representation of the commutator. In particular, by evaluating the integral over the cosine, we find 
\begin{flalign}
\langle {\bar \phi}|[{\phi}(x),{\phi}(0)]|{\bar \phi}\rangle&=\frac{1}{2\pi \ri} \delta({\vec x }^{\, 2}-t^2)-\frac{\lambda {\bar \phi}^{\, 2}}{16\pi  \ri \,|\vec x|}\, \theta(t-|\vec x|)\int_{\frac{t-|\vec x|}{2}}^\frac{t+|\vec x|}{2}  {\rm d} z^0 \nonumber\\&=\frac{1}{4\pi \ri |\vec x| } \delta(|\vec x|-t)-\frac{\lambda {\bar \phi}^{\, 2}}{16\pi  \ri}\, \theta(t-|\vec x|)~, \qquad t>0.
\label{eq:firstorderQuartic}
\end{flalign}
Notice that, because we focus on $t>0$, we have a theta function that provides support only on the future light cone. By consistency, in the last step we expanded the delta function and considered only the part that has support in the same region.\footnote{To evaluate the integral \eqref{eq:firstorderQuartic}, we write the integration range over the angle as 
\begin{equation}
 \int_{0}^tdz^0\int_{-\infty}^\infty d\cos\vartheta\left\{\theta(\cos\vartheta-1)-\theta(\cos\vartheta+1)\right\}\delta(\cos\vartheta-...)= \theta(t-|\vec{x}|)\int_{0}^tdz^0 \left\{\theta\left(z_0-\frac{x+t}{2}\right)-\theta\left(z_0-\frac{x-t}{2}\right)\right\}.
\end{equation}
In the second step, we integrated the delta function over the angle, and then reduced the argument of the heaviside theta $\theta$ using the fact that $0\leq |\vec{x}|\leq t$. The overall heaviside theta enforces that the integral has no support outside the light-cone of Minkowski.
}

As we mentioned, interactions made the commutator non-vanishing for time-like separations. This correction is equivalent to what one would obtain in a free massive scalar field theory by treating the mass term as a perturbation and including it in the interaction Hamiltonian. 
This follows from the fact that correlation functions evaluated on the coherent state $|\bar{\phi}\rangle$ are equivalent to those obtained by studying fluctuations around the corresponding background $\bar{\phi}$. Introducing the field decomposition $\phi_0(x,t)={\bar \phi}+\varphi_0(x,t)$, the quartic potential reorganizes into quadratic, cubic, and quartic terms. The quadratic term defines a mass for the fluctuation field, given by $m^2=\lambda {\bar \phi}^{\, 2}/2$. By treating this mass as a perturbation, the first order correction to the commutator of fluctuations $\varphi_0(x,t)$ precisely matches the result found in Eq.~\eqref{eq:firstorderQuartic}. 

\section{Derivative interactions: microcausality in $P(X)$ theories}
\label{sec3}

In the example above, interactions modify the commutator minimally. The light-cone is basically left unchanged, it is just ``filled up", in the sense the support to the commutator extends inside the light-cone in the presence of interactions.

As discussed in the introduction, however, derivatively coupled theories are expected to modify the causal structure of the free theory more dramatically. Gravity is, obviously, one of these theories, to which we will turn in an upcoming work. 
In this section we take on an easier piece, that of a $P(X)$ theory of the type
\begin{equation}
     \mathcal{L}=-\frac{1}{2}\partial_\mu \phi \partial^\mu \phi+\frac{\alpha}{4} \left(\partial_\mu \phi \partial^\mu \phi\right)^2\, .
     \label{eq:quarticderivative}
\end{equation}
This Lagrangian is Lorentz-invariant, so the argument sketched in the introduction applies, that commutators at space-like separations are unitarily related to commutators evaluated at the same time. The vanishing of the latter should imply the vanishing of the former. 
However, this reasoning seems to fail for the above theory, especially when $\alpha<0$ and  fluctuations around certain Lorentz-breaking backgrounds feature superluminal propagation. 

Our perturbative approach appears to accommodate both perspectives. At each order in perturbation theory the commutator remains formally confined on the Minkowski light-cone. However, its expression contains derivatives of delta functions,  which we show  are associated to a perturbative change of the causal structure.

As this theory contains time derivatives in the interaction terms, the interaction picture formalism should be applied with some care.  
First we identify the conjugate momentum, which reads
\begin{flalign}
&\pi=
\dot{\phi}+\frac{\alpha}{2}\left[2\dot{\phi}^3-\dot{\phi}\left(\partial_i \phi \partial^i\phi\right)-\left(\partial_i \phi \partial^i\phi\right) \dot{\phi}\right]. 
\label{eq:conjugateM}
\end{flalign}
As discussed in Section~\ref{Sect:interactionPict}, it is essential that we remove the field velocity in favor of the canonical momentum in the full Hamiltonian, before applying the interaction picture formalism. For this, we have to invert the relation \eqref{eq:conjugateM}, which cannot be done exactly. However, since we are interested in the first-order correction in the coupling, we invert it perturbatively in the number of fields and focus on the leading-order correction.\footnote{In particular, the first order solution to \eqref{eq:conjugateM}  reads
\begin{flalign}
    \dot{\phi}=\pi-\frac{\alpha}{2}\left[2\pi^3-\left(\partial_i \phi \partial^i\phi\right) \pi-\pi\left(\partial_i \phi \partial^i\phi\right) \right]+\ldots \nonumber
\end{flalign} }
Once the relation has been inverted, we plug it back into the full Hamiltonian and identify the following free and interaction parts
\begin{flalign}
    &\mathcal{H}_0=\frac{1}{2}\dot{\phi}_0^2+\frac{1}{2}\partial_i \phi_0 \partial^i \phi_0,
    \label{eq:freeHPX}\\ &\mathcal{H}_\text{I}=-\frac{\alpha}{4} \left({\dot{\phi}}_0^4+2\dot{\phi}_0\left(\partial_i \phi_0 \partial^i\phi_0\right)\dot{\phi}_0-2\left\{\left(\partial_i \phi_0 \partial^i\phi_0\right),\dot{\phi}_0^2\right\}+(\partial_i \phi_0 \partial^i \phi_0)^2\right),
    \label{eq:IntH}
\end{flalign}
where the curly brackets indicate the anti-commutator. In deriving the above Hamiltonians, the field and conjugate momentum are treated as non-commuting variables. 
We also applied the relation $\pi_0(x)=\dot{\phi}_0(x)$.

By using~\eqref{eq:IntH}  into Eq.~\eqref{eq:commutatorInt1}, we obtain the following expression for the commutator of the Heisenberg fields
\begin{flalign}
[\phi(x),\phi(0)]=\nonumber &
\, \Delta(x)+i\alpha \int_{0}^{t} d^4 z \biggr[\mathcal{A}(z){\partial_{0} \Delta(z-x)}\partial_{0} \Delta(z)+\mathcal{B}_{ij}(z) {\partial_i \Delta(z-x)} {\partial_j \Delta(z)}\nonumber\\&+\mathcal{C}_{i}(z)\biggr({\partial_{0}}\Delta(z-x){\partial_{i}}\Delta(z)+{\partial_{i}}\Delta(z-x){\partial_{0}} \Delta(z)\biggr) \biggr],
\label{eq:commutator1}
\end{flalign}
where all partial derivatives acting on the $\Delta$ correlators  are with respect to the $z$ variable and $ \mathcal{A}(z),  \mathcal{B}_{ik} (z) $ and $ \mathcal{C}_i (z)$ are products of local free fields operators evaluated at $z$, 
\begin{flalign}
\label{eq:coeff1}
    \mathcal{A}&=-3\dot{\phi}^2_0 + \partial_i \phi_0 \partial^i \phi_0\, , \\
    \mathcal{B}_{ik}&=\left(\dot{\phi}^2_0-\partial_i \phi_0 \partial^i \phi_0\right)\delta_{ik}-2\partial_i \phi_0 \partial_k \phi_0 \label{eq:coeff2}\, , \\ 
    \mathcal{C}_i&=\big\{\partial_i\phi_0, \dot{\phi}_0\big\}\label{eq:coeff3}\, .
\end{flalign}

\begin{figure} 
	\centering
	\includegraphics[clip,scale=0.19, trim=1cm 0cm 1cm 0.5cm,]{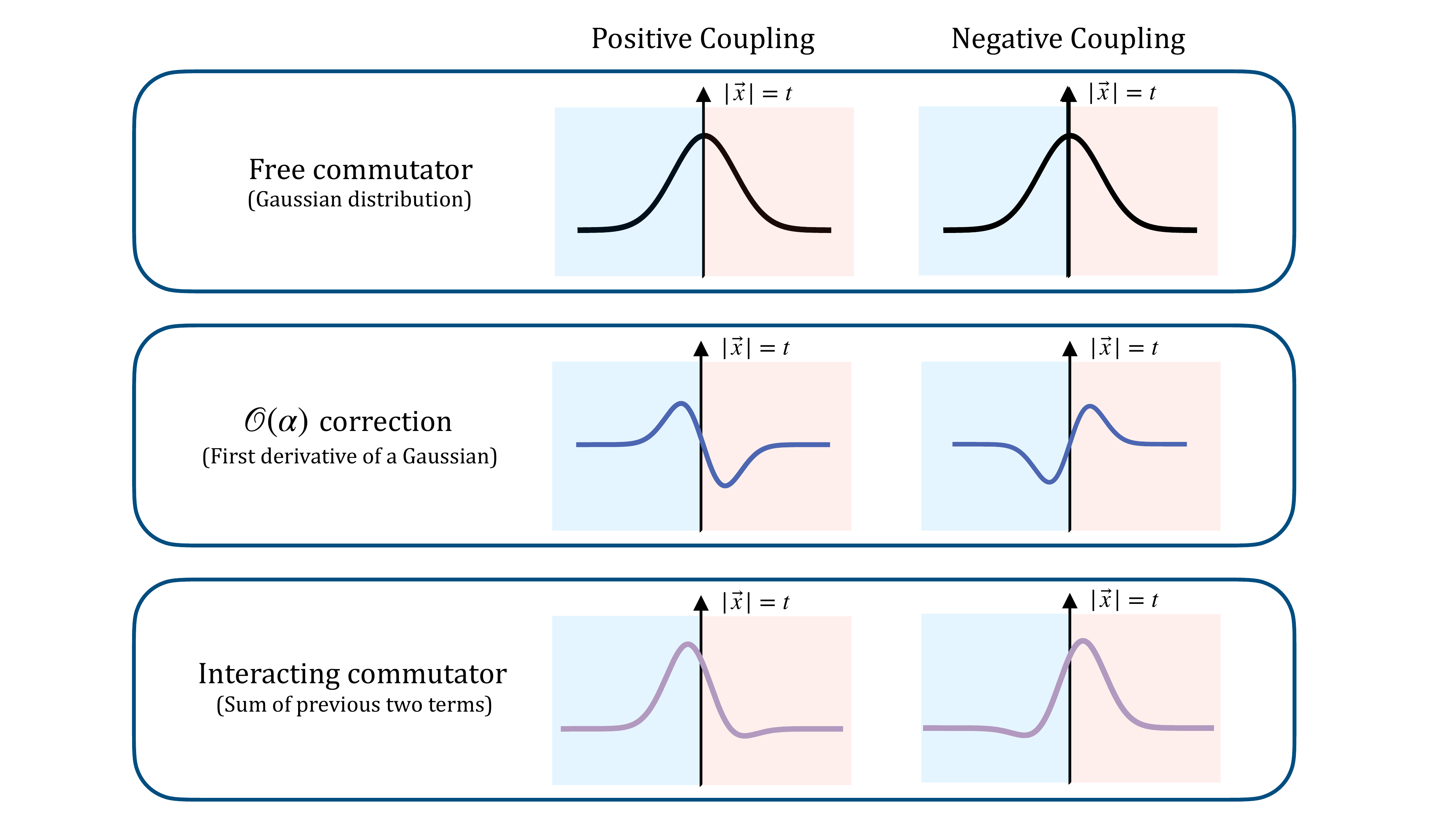} 
\caption{Pictorial representation of the  commutator in a $P(X)$ theory. Delta functions and their derivatives are regularized using Gaussians with finite variance. The blue and red shaded regions denote the interior and exterior of the light-cone, respectively. The free commutator is modeled as a Gaussian sharply peaked on the light-cone. In contrast, the first-order correction appears as the derivative of a Gaussian—antisymmetric and vanishing exactly on the light-cone. Its interference with the free commutator depends on the sign of the coupling: for positive coupling, the correction enhances the free Gaussian inside the light-cone and suppresses it outside; for negative coupling, the opposite occurs.  In the limit of vanishing variance, the sum approximate a delta-function supported on the peak of the new distribution.}
	\label{fig:deltaInterf}
\end{figure}

Eq.~\eqref{eq:commutator1} is not very transparent in its present form.  However,  we immediately observe that it contains terms involving derivatives of delta functions, as it contains derivatives of the free commutator~\eqref{eq:commutatorFree}. By studying its expectation value over a Lorentz breaking state, in the next section we show that  this corrections are responsible for modifying the light-cone of the free theory. 

There is a straightforward, albeit perhaps naive, way to understand this mechanism in terms of a regularized version of the delta function and its derivatives. 
One might notice that the correction term in~\eqref{eq:commutator1} is the same for well behaved ($\alpha>0$) and pathological ($\alpha <0$) theories. Since it is multipied by $\alpha$ it just comes with different signs.  Now, the delta derivative contained in the correction term is odd across the light-cone, and can thus move the peak of the commutator inside or outside the light-cone depending on the sign of $\alpha$ (Fig.~\ref{fig:deltaInterf}). 

\subsection{A Lorentz-breaking coherent state}
\label{sec:ExpCOHOsc}

We now want to write a state $\mu$ which is the quantum mechanical equivalent of the classical field configuration $\phi = \mu t$. These ``spontaneous symmetry probing"  states are special, in that their time evolution in field-space follows a symmetry direction for the theory~\cite{Nicolis:2011pv}. It can be shown that these states are protected against quantum \emph{depletion}, although they should be constructed with some care. In particular they are non-Gaussian states~\cite{Berezhiani:2025tkp},  because the Lagrangian contains interaction terms.

Here we content ourselves with a Gaussian approximation of these states,
\begin{equation}
|\mu\rangle=e^{\ri {f}}|0\rangle,\qquad \text{with} \qquad {f}=\mu\left(1+\alpha{\mu^2}\right) \int d^3 x  \,\phi(\vec{x},0)\, .
\label{eq:state}
\end{equation}
One can check that, on sufficiently short time scales,  their expectation values indeed match the classical behavior, 
\begin{equation}
    \langle \mu|\phi(x)|\mu\rangle=\mu t,\qquad \langle \mu|\pi(x)|\mu\rangle=\mu\left(1+\alpha{\mu^2}\right)\, .
    \label{eq:lorentzb}
\end{equation}

Using the state defined above, we can consider the expectation value of the commutator \eqref{eq:commutator1} and retain only the leading correction in the coupling. According to the discussion of Appendix~\ref{sec:COH}, we find the following relations 
\begin{flalign}
&\langle \mu|\dot{\phi}^2_0|\mu\rangle=\mu^2+\langle 0|\dot{\phi}_0^2|0\rangle , \\ & \langle \mu|\partial_i\phi_0\partial^i \phi_0|\mu\rangle= \langle 0|\partial_i{\phi}_0\partial^i \phi_0|0\rangle, \\
&\langle \mu|\partial_i\phi_0\dot{\phi}_0|\mu\rangle=0
 \label{eq:commutatorC}
\end{flalign}
to leading order in $\alpha$.
We use this to evaluate the expectation values of $\mathcal{A}$, $\mathcal{B}_{ij}$ and $\mathcal{C}_i$ over $|\mu\rangle$.  
With these,
 the expectation value of the commutator reads
\begin{flalign}
\langle \mu|[\phi(x),\phi(0)]|\mu\rangle= 
\Delta(x)+{\ri\alpha \mu^2}\int_{0}^{t} d^4 z \biggr[-3\partial_{0} \Delta(z-x)\partial_{0} \Delta(z)+ \partial_i\Delta(z-x)\partial^i \Delta(z) \biggr]+\ldots
\label{eq:PX1order}
\end{flalign}
Notice that the above correction depends on the effective dimensionless coupling $\alpha \mu^2$.
As in the case of the coherent state in the quartic theory, we remove the purely divergent terms $\langle 0|\phi_0^2|0\rangle$ and $\langle 0|\partial_i\phi_0\partial^i \phi_0|0\rangle$ by renormalizing the theory.  These can be eliminated by introducing the standard wavefunction counterterm in the interaction Hamiltonian~\eqref{eq:IntH}, that exactly cancels them out.

As mentioned, the above terms clearly give rise to derivatives of the delta function.  However, rather than using the expression for the free commutator~\eqref{eq:commutatorFree} and working directly in real space, it is more prudent to take a detour through momentum space to evaluate the integral.  A step-by-step evaluation of the integral is provided in Appendix~\ref{app:EvaluationComm},  yielding the result
\begin{align}
\label{eq:PX1orderfraction}
 \langle \mu|[\phi(x),\phi(0)]|\mu\rangle& = \frac{\text{sign}(t)}{2\pi \ri}\delta(\vec x^{\, 2}-t^2)\\[2mm] +&\alpha{\mu^2}\left[\frac{1}{2\pi^2}\left(\frac{1}{t_-^2-\vec x^{\, 2}}-\frac{1}{t_+^2-\vec x^{\, 2}}\right)\nonumber -\frac{ t^2}{2 \pi ^2}\left(\frac{1}{(t_-^2-\vec x^{\, 2})^2}-\frac{1}{(t_+^2-\vec x^{\, 2})^2}\right)\right]\, ,
\end{align}
where $t_\pm=t(1\pm\ri \epsilon)$.  In particular,  the momentum integrals exhibit oscillatory behavior, and to ensure convergence, different terms require distinct prescriptions for analytic continuation into imaginary time.

By identifying the discontinuities in Eq.~\eqref{eq:PX1orderfraction},  this expression can be further reduced to the final form
\begin{flalign}
\label{eq:commutatorDerivative}
\langle \mu|[\phi({x}),\phi(0)]|\mu\rangle
=&\, \frac{\text{sign}(t)}{2\pi\ri}\left(1+\alpha{\mu^2} |\vec x|\,\frac{\partial}{\partial |\vec x|} \right)\delta(\vec x^{\, 2}-t^2)\\
=&\, \frac{\text{sign}(t)}{2\pi\ri}\left[\left(1-2\alpha\mu^2\right)\delta(\vec x^{\, 2}-t^2)+2\alpha{\mu^2}t^2 \,\delta'(\vec x^{\, 2}-t^2)\right]\, .
\label{eq:commutatorDerivative2}
\end{flalign}

As anticipated, the first-order correction to the free propagator takes the form of a derivative of a delta function with support on the light-cone of the free theory.  In the remainder of this work,  we show that this type of corrections is associated with a modification of the light-cone structure of the free theory.  We can already see this by noting that,  in Eq.~\eqref{eq:commutatorDerivative2},   the coefficient of the derivative of the delta function matches the expansion given in Eq.~\eqref{deltader_intro} in the introduction.  
In the next section,  we show that it corresponds to the first-order term in the Taylor expansion of a commutator supported on a modified light cone,  expanded around the Minkowski one. 
The first line of Eq.\eqref{eq:commutatorDerivative} provides a compact expression for the commutator, which we will use in Section \ref{sec_extended}.

Notice that while the derivatives of the delta function are generically present also in the general operatorial commutator~\eqref{eq:commutator1}, they become ``activated" only on certain states. Whether or not they survive also in the \emph{expectation value} of the commutator, is a matter of balance between the time space derivative terms appearing in Eq.~\eqref{eq:PX1order}. A notable counter example is that of a thermal state for the $P(X)$ at zero chemical potential in Sec. \ref{eq:thermalPX}. This state breaks boosts but does not ``activate" the delta derivative terms and thus preserves the Lorentz invariant light-cone. 

 Finally, we note that the first-order correction given in~\eqref{eq:commutatorDerivative2} appears to exhibit a secular behavior. Secular terms can compromise key properties of correlation functions, such as the boundedness and positivity of their spectral representation. These issues typically arise when computations are performed beyond the semiclassical regime, where loop corrections become relevant. In such cases, it is necessary to adopt appropriate truncation schemes that preserve these properties~\cite{kraichnan1961dynamics,Mihaila:2000sr}. In our case, however, a resummation of the semiclassical result is known and corresponds to a Dirac delta supported on the semiclassical sound-cone of the theory of fluctuations around the background~\eqref{eq:lorentzb}. Moreover, reading the secular behavior directly from \eqref{eq:commutatorDerivative2} is not straightforward due to the presence of the derivative of the delta function. In Section~\ref{sec_extended}, however, we show that the approximation remains valid at all times, as long as $\alpha \mu^2\ll 1$. We prove this by using linear response theory in the presence of an extended Gaussian source.  

\subsection{Expanding in small fluctuations around the background field} \label{sec_fluctuations}

The expectation value of the commutator on the coherent state $|\mu\rangle$, as given by Eq.~\eqref{eq:commutatorDerivative}, has to match the commutator obtained by considering fluctuations $\varphi(x)$ around the classical solution $\phi_0(t)=\mu t$. In the same way of the quartic potential, at this order, the connected part of correlation functions of the fundamental field $\phi(x)$ evaluated on a coherent state describing a given classical configuration coincides with the connected part of the corresponding correlation functions expressed in terms of fluctuations $\varphi(x)$ around the same configuration.\footnote{However, this equivalence holds only at the tree-level. At higher orders, the coherent state encodes the full quantum evolution of the system, whereas the classical solution used in semiclassical dynamics does not. For instance, if we push the computation to higher loop orders, we would realize that the background we have to consider in the semiclassical expansion to capture the dynamics of the coherent state is the one-point function of the field operator itself, and not the classical solution.
The difference between the two backgrounds is described by the \textit{quantum backreaction of modes}, and to incorporate it correctly it requires studying the quantum dynamics of the background and the fluctuation together, in a coupled system of equation of motions. See Ref. \cite{Berezhiani:2021gph}.}

To show this, we introduce the following field redefinition
\begin{equation}
    \phi(x)=\mu t+\varphi(x),
\end{equation}
and apply it to the theory \eqref{eq:quarticderivative}. The Lagrangian of the fluctuation field $\varphi$ reads
\begin{flalign} \label{lagrangian}
 \mathcal{L}=\frac{1}{2}\left(1+3\alpha{\mu^2}\right)\dot{\varphi}^2-\frac{1}{2}&\left(1+\alpha{\mu^2}\right)\partial_i\varphi \partial^i\varphi+\ldots
\end{flalign}
The commutator for the field $\varphi(x)$ deviates from the Lorentz light cone already at the level of this quadratic Lagrangian. The easiest way to compute is to rescale the spatial coordinates $\vec x \rightarrow c_s \vec x$ to make the theory relativistic and the field $\varphi \rightarrow \tilde \varphi$ to make it canonically normalized. In these coordinates, the commutator for $\tilde \varphi$ is clearly the standard one~\eqref{eq:commutatorFree}. This trick gives 
\begin{equation}
 [{\varphi}(x),{\varphi}(0)]=\left({1+3\alpha{\mu^2}}\right)^{-1/2}\left({1+\alpha{\mu^2}}\right)^{-1/2}\frac{\text{sign}(t)}{2\ri \pi }\delta\left(\vec x^{\, 2}-c_s^2t^2\right)
 \label{eq:commutatorvarphi}
\end{equation}
where we have introduced the speed of sound 
\begin{equation}
    c_s^2=\frac{1+\alpha{\mu^2}}{1+3\alpha{\mu^2}}\, .
    \label{eq:SS}
\end{equation}

In order to compare Eq.~\eqref{eq:commutatorvarphi} to the expectation value of the commutator of the fundamental field over the coherent state, we perform a $\alpha\mu^2$ expansion up to the first order.
The expansion of prefactors is straightforward, while for the delta-function we can use Eq.~\eqref{deltader_intro} 
and $c_s^2\simeq 1-2\alpha{\mu^2}$. We obtain the following expansion of the commutator of $\varphi(x)$
\begin{flalign}
 [{\varphi}(x),{\varphi}(0)]= \frac{\text{sign}(t)}{2\pi\ri}\left[\left(1-2\alpha\mu^2\right)\delta(\vec x^{\, 2}-t^2)+2t^2\alpha{\mu^2}\,\delta'(\vec x^{\, 2}-t^2)\right]+\ldots
\end{flalign}
We appreciate that it matches the expectation value of the commutator \eqref{eq:commutatorDerivative2}, which we evaluated in the previous section. Furthermore, we find that the additional term proportional to the delta function arises from the modification of the canonical normalization of the field induced by the coherent state,  leading to the prefactors appearing in Eq.~\eqref{eq:commutatorvarphi}.
\subsection{Thermal averaged commutator in \textit{P(X)}} \label{eq:thermalPX}

Finally,  we examine if all states that spontaneously break Lorentz boosts provide a commutator that contains derivative of delta-functions.  Here,  we show that this is not the case by studying the finite temperature free vacuum state of the theory $|0\rangle_\beta$,  with $\beta=1/T$. 

In particular,  we evaluate the ensemble average of the interacting commutator in the thermal configuration. 
We define the ensemble average of a general operator $\mathcal{O}$ as
\begin{equation}
    \langle \mathcal{O}\rangle_\beta=
    \frac{\text{Tr}~\left(e^{\beta H}\mathcal{O}\right)}{\text{Tr} ~e^{\beta H}},
\end{equation}
The function $H$ is the full Hamiltonian of the $P(X)$ theory, implying we consider a configuration of the scalar field which in thermal equilibrium and with zero chemical potential.

The thermal averaged commutator~\eqref{eq:commutator1} reads 
\begin{flalign}
\langle [\phi(x),\phi(0)]\rangle_\beta=\nonumber &
\Delta(x)+{\ri}\alpha\int_{0}^{t} d^4 z  \biggr\{\langle\mathcal{A}(z)\rangle_\beta{\partial_{0} \Delta(z-x)}\partial_{0} \Delta(z)+\langle \mathcal{B}_{ij}(z)\rangle_\beta {\partial_i \Delta(z-x)} {\partial_j \Delta(z)}\nonumber\\&+\langle\mathcal{C}_{i}(z)\rangle_\beta\biggr({\partial_{0}}\Delta(z-x){\partial_{i}}\Delta(z)+{\partial_{i}}\Delta(z-x){\partial_{0}} \Delta(z)\biggr) \biggr\},
\label{eq:commutator2}
\end{flalign}
where $\langle\mathcal{A}(z)\rangle_\beta$, $\langle \mathcal{B}_{ij}(z)\rangle_\beta$ and $\langle\mathcal{C}_{i}(z)\rangle_\beta$ are the thermal averages of the operators \eqref{eq:coeff1}, \eqref{eq:coeff2} and \eqref{eq:coeff3}.

In order to evaluate these thermal averaged correlation functions,  we need to first find 
$\langle \dot{\phi}_0^2\rangle_\beta$ and $\langle \partial_i{\phi}_0\partial_j \phi_0\rangle_\beta$. Since we require them at leading order in the coupling, we approximate ${H}\simeq {H}_0$ and evaluate the thermal averages as in the free theory. 
To do this, we use the real-time formalism \cite{Das:1997gg}.~\footnote{Given that we are working with an equilibrium configuration, the real-time and imaginary-time formalisms coincides.  However, the former explicitly retains the notion of time, and we do not have to perform a analytic continuation to recover standard Minkowski time from its imaginary counterpart.} With this approach, the perturbative time evolution of correlation functions is described in terms of a time-ordered $G_{++}(x-y)$, an anti-time-ordered $G_{--}(x-y)$, and two out-of-time-ordering propagators $G_{+ -}(x-y)$ and $G_{- +}(x-y)$.  Since the operators appearing in \eqref{eq:commutator2} are evaluated at the same time,  the final result is unaffected by the specific choice of the propagator.

Therefore, we introduce the following thermal representation of the quadratic correlation function of the field operator
\begin{equation}
    \langle \phi(x)\phi(y)\rangle_\beta=\ri G_{-+}(x-y)=\int \frac{d^3{k}}{(2\pi)^3 2|\vec{k}|}\left\{\left(1+n_B({|\vec{k}|})\right)e^{-\ri |\vec{k}|(t-t')}+n_B(|\vec{k}|)e^{\ri|\vec{k}|(t-t')}\right\}e^{\ri \vec{k}\cdot (\vec{{x}}-\vec{{y}})},
\end{equation}
where the time labels satisfy the relation  $t'>t$ and where $n_B(x)=({e^{\beta x}-1})^{-1}$. Then, we can use this expression to evaluate the following correlation functions
\begin{flalign}
     &\langle \{\partial_i\phi_0, \dot{\phi}_0\}\rangle_\beta=0\\&
     \langle \dot{ \phi}^2_0\rangle_\beta=\lim_{z_1\to z_2}\frac{\partial}{\partial {z^0_1}}\frac{\partial}{\partial {z^0_2}}\langle  \phi(z_1) \phi(z_2)\rangle_\beta=\int \frac{d ^3{k}}{2(2\pi)^3 }|\vec{k}|\left\{1+2n_B(|\vec{k}|)\right\}\\&
      \langle \partial_i \phi_0\partial_j \phi_0\rangle_\beta=\lim_{z_1\to z_2}\frac{\partial}{\partial {z^i_1}}\frac{\partial}{\partial {z^j_1}}\langle  \phi(z_1) \phi(z_2)\rangle_\beta=\delta_{ij}\int \frac{d ^3{k}}{2(2\pi)^3 }|\vec{k}|\left\{1+2n_B(|\vec{k}|)\right\}
\end{flalign}
One can show that using any of the other propagators provides the same result.
By plugging these relations back into the thermal averaged commutator of Eq.~\eqref{eq:commutator2}, we obtain the following expression
\begin{flalign}
\langle [\phi(x),\phi(0)]\rangle_\beta=\nonumber &
\Delta(x)-i \alpha\int\frac{d^3k}{(2\pi)^3}|\vec{k}|\left(1+2n_B(|\vec{k}|)\right)\times\\&\int_{0}^{t} d^4z \biggr\{\partial_{0} \Delta(z-x)\partial_{0} \Delta(z)- \partial_i\Delta(z-x)\partial^i \Delta(z)\biggr\}.
\label{eq:commutator3}
\end{flalign}
We can already compare it to the expectation value of the same commutator over the Lorentz-breaking coherent state, provided by Eq.~\eqref{eq:commutatorC}. 
What we see is that there is a different relative factor between the correlation functions involving time and space derivatives. In this case, we have $-1$.  Following App.~\ref{app:EvaluationComm},  the derivative of delta functions cancels out between the two different contributions providing the following final result
\begin{equation}
    \langle [\phi(x),\phi(0)]\rangle_\beta=\frac{\text{sign}(t)}{2\pi\ri}\left(1-\alpha\frac{\pi^2 }{15 }{T^4}-\alpha\frac{\Lambda_\text{UV}^4}{8 \pi^2}\right)\delta(\vec{x}^{\,2}-t^2).
\end{equation}
Here, $\Lambda_\text{UV}$ is the ultraviolet cutoff that we had to introduce to regulate the momentum integral appearing in Eq.~\eqref{eq:commutator3}.
This term is temperature-independent, as expected, and is removed by introducing the same wave-function renormalization used in the case of the coherent state \eqref{eq:state}.

We conclude that thermal configurations with zero chemical potential do not change the light-cone structure of the theory.

\section{Linear response to an extended source}\label{sec_extended}

As we have seen from  Eq.~\eqref{eq:commutatorvarphi}, the free commutator of the field fluctuations $\varphi$,  on the Lorentz-breaking state $|\mu\rangle$,  has support on the ``sound-cone", centered at $\vec x^{\, 2} = c^2_s t^2$. On the other hand, when we evaluate the commutator of the canonical field $\phi$ in perturbation theory, we obtain Eq.~\eqref{eq:commutatorDerivative2} which, at least formally, has still support on the standard lighcone at  $\vec x^{\, 2}=t^2$. From the leading terms we have computed, it is evident that the result corresponds to a Taylor expansion of a delta function centered on the sound-cone, expanded around the standard light-cone, as illustrated in Eq.~\eqref{deltader_intro}.
One might wonder whether an entire resummation is needed to see the effect. 
In this section we show that a modified causal structure is already implied by the first ($\delta '$) term of the perturbative expansion. 

\begin{figure} 
	\centering
	\includegraphics[scale=0.5]{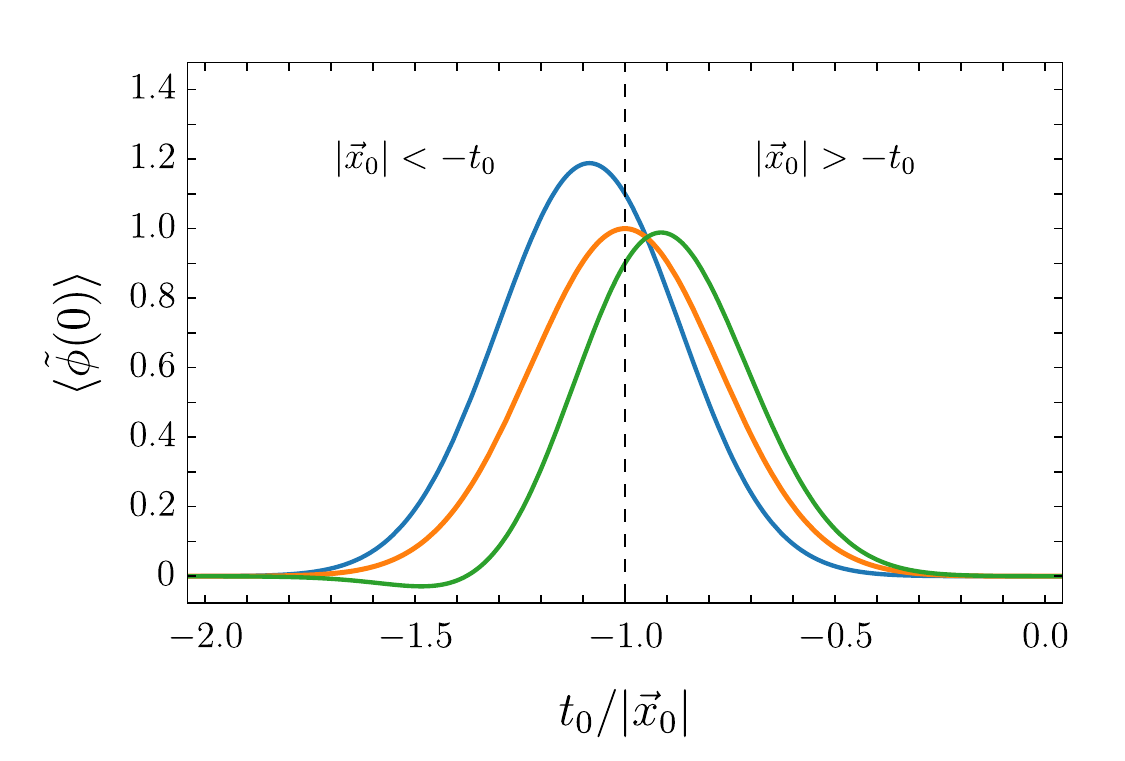} 
\caption{The plot shows the response of the expectation value of the field over the coherent state $|\mu\rangle$ to the external Gaussian source \eqref{eq:source}, evaluated at the origin. The response is given in function of the time $t_0$, normalized to the value $|\vec{x}_0|$. On the y-axis, we plot the canonically rescaled field $\tilde{\phi}(x)=J^{-1}_0\sqrt{2\pi}\sigma |\vec{x}_0| \phi(x) $, and we fixed $|\vec{x}_0|^2=20 \sigma^2 $. 
The Red, Blue, and Purple curves are the response for a negative, positive, and null coupling respectively. When the theory is free ($\alpha=0)$, the response is maximized for a Gaussian source centered on the past light cone of Minkowski. If the coupling is positive/negative, the response is maximized for a source centered inside/outside the past light cone. We conclude that for $\alpha<0$ we have superluminal propagations. }
	\label{fig:response}
\end{figure}

In order to do this, we apply linear response theory to an extended source $J(x)$.\footnote{We thank A. Nicolis for suggesting this computation} 
It proves convenient to calculate the response of the field at the origin and center the source around an event $x_0$ in the past, 
\begin{equation}
    \langle \mu| \phi(0)|\mu \rangle=
    \langle \mu|\phi(0)|\mu\rangle_{J=0}-\ri \int_{-\infty}^{0} d^4x~J(x)\langle \mu|[\phi(x),\phi(0)] |\mu\rangle\, ,
    \label{eq:response}
\end{equation}
with
\begin{equation}
    J(x)=\frac{J_0}{\pi^2\sigma^4} \text{exp}\left(-\frac{(\vec{{x}}-\vec{{x}}_{0})^2+{(t-t_0)^2}}{{\sigma^2}}\right).
    \label{eq:source}
\end{equation}
The above corresponds to a gaussian source centered in $\vec{{x}}=\vec{{x}}_0$ and peaked at the time $t=t_0$. Of course this expression is not Lorentz invariance and appears ``spacetime spherically symmetric" only in one frame. We obtain
\begin{flalign}
    \langle \mu| \phi(0)|\mu \rangle =&\frac{J_0}{\pi^2\sigma^4}\int_{-\infty}^0 d^4 x ~e^{-\frac{(\vec{{x}}-\vec{{x}}_{0})+(t-t_0)^2}{\sigma^2}} \left(1+\alpha{\mu^2}|\vec{x}|\frac{\partial}{\partial|\vec{x}|}\right)\delta(\vec{x}^{\,2}-t^2))\nonumber\\[2mm]=& \, \frac{J_0}{\pi^2\sigma^4}\int_0^\infty  |\vec{x}| d |\vec{x}| d\Omega\left\{\left(1-\alpha{\mu^2}+2\alpha{\mu^2}\frac{|\vec{x}|^{2}-|\vec{x}| |\vec{{x}}_0| \cos\vartheta}{\sigma^2}\right)\right\}e^{-\frac{(\vec{x}-\vec{x_0})^2+(|\vec{x}|+t_0)^2}{\sigma^2}},
\end{flalign}
where we exploit the fact that $\langle \mu|\phi(0)|\mu\rangle_{J=0}=0$.
The overall minus in the first line is compensated by the fact that the time $t<0$ is always negative within the integration contour, and the commutator flips the sign. The second line is obtained by performing an integration by part on the internal $|\vec{x}|$-derivative and decomposing the delta function according to Eq.~\eqref{eq:deltarelatin}. Then, we
integrate over the internal time $t$, and we drop $\delta(t-|\vec{x}|)$ since it has no support in the overlap of the integration regions $0<|\vec{x}|<\infty$ and $-\infty<t<0$. 

The above integral can be evaluated exactly in terms of error functions.   In Fig.~\ref{fig:response}, we plot the exact solution for a positive, negative, and null value of the effective coupling $\alpha \mu^2$.   To write the final result in a more manageable form, we expand it in the far-source limit $\sigma/ |\vec{x}_0| \to 0$, obtaining
\begin{equation}
  \langle \mu|\phi(0)|\mu\rangle_J= \frac{J_0}{\sqrt{2\pi }\sigma |\vec{x}_0|} \left[1+\frac{\alpha\mu^2}{2}\left(1-\frac{|\vec{x}_0|^2-t_0^2}{\sigma^2}\right)\right] e^{-\frac{ (t_0+|\vec{x}_0|)^2}{2\sigma^2}}\, .
  \label{eq:responseexpande}
\end{equation}
We only included the first correction in this expansion, which shows the modification of the theory's causal response induced by the derivative of the delta function.

Now we look for the values of $\vec{x}_0$ and $t_0$ that maximize the field response at the origin~\eqref{eq:responseexpande}. At leading order in the coupling, the maximum occurs at~\footnote{A second maximum appears at $|\vec{x}_0|=0$. However, this contribution is spurious, as we are considering the regime where the source is located far from the origin. To avoid taking it into account,  we find the maxima by varying Eq.~\eqref{eq:responseexpande} with respect to $t_0$. }
\begin{equation}
    |\vec{x}_0|=-t_0\left(1-\alpha{\mu^2}\right).
    \label{eq:Maximum}
\end{equation}
Let us interpret this result. In the absence of interactions ($\alpha = 0$), an observer at the origin experiences the maximum field response when the Gaussian source is centered on their past light cone. This aligns with the expectation that the strongest response originates from the spacetime point where the source is maximized, with the signal then propagating to the observer. In the free theory, this propagation occurs exactly at the speed of light, as dictated by the free commutator Eq.~\eqref{eq:commutatorFree}. In the presence of interactions, spacetime points contributing to the maximum response are affected. For a positive coupling ($\alpha > 0$), the response is maximized when the peak of the source lies \textit{inside} the past light cone, effectively slowing down the propagation to subluminal speeds. Conversely, for a negative coupling ($\alpha < 0$), the maximum shifts \textit{outside} the Minkowski light cone, meaning the quartic term modifies the signal propagation speed to superluminal values (Fig.~\ref{fig:response}).
The key point to emphasize is that, for any value of $\mu$, the maximum lies outside the light-cone when $\alpha<0$. This indicates that signatures of superluminality emerge within the regime of validity of the effective field theory.

Finally, let us note that if we had used the resummed semiclassical result \eqref{eq:commutatorvarphi}, the response would have been maximized for a Gaussian source centered at $ |\vec{x}|=-c_s t$. Expanding the semiclassical sound speed \eqref{eq:SS} to first order in $\alpha \mu^2$ precisely reproduces the result \eqref{eq:Maximum}, further validating our analysis.  Moreover, this indicates that, as long as $\alpha \mu^2\ll 1$, our result provides a good approximation of the semiclassical one, where all classical non-linearities of the theory are resummed.

\section{Conclusions}
\label{sec5}
In this paper, we have studied operator-valued commutators in scalar field theories, with particular interest in those that exhibit a propagation speed different from the speed of light on certain backgrounds—such as the  $P(X)$ of Eq.~\eqref{eq:quarticderivative}.  Before being evaluated on any state, the interacting commutator should maintain a Lorentz invariance appearance---the $P(X)$ is Lorentz invariant---but enough flexibility to trace different lighcone structures when evaluated on different states. We have shown that this is achieved by the derivatives of the delta functions with support on the standard Minkowski light-cone, which appears in the perturbative correction to the commutator~\eqref{eq:commutator1}. 

While perturbative calculations manage to save appearances in regard to Lorentz invariance, we know, on the other hand, that  
when the interaction term of the $P(X)$ has the ``wrong sign" ($\alpha<0$) causality violations \emph{are} at play.  Small fluctuations around a Lorentz-breaking background become superluminal (see Sec.~\ref{sec_fluctuations}), and even just the first perturbative correction to the commutator~\eqref{eq:commutator1} clearly leads to a superluminal linear response in the case of extended sources, as shown in Sec.~\ref{sec_extended}. At the same time, when $\alpha<0$, the theory cannot be consistently completed in the UV~\cite{Adams:2006sv}. Still, regardless of the sign of $\alpha$, $P(X)$ is a low-energy relativistic theory. It would be nice to understand precisely what goes wrong with the standard argument that Lorentz invariance implies microcausality. 

Let us review this argument. If boosts are implemented unitarily on the Hilbert space of the theory, commutators outside the lightcone are unitarily equivalent to commutators on equal time surfaces. The latter should vanish by construction, as a standard quantization condition. So also commutators at spacelike separation must vanish.
Since effective field theories come with a built-in energy cutoff, one might argue that their spacetime description is inevitably blurred, coarse grained. And that boosting an operator up to the lightcone would require a spacetime resolution well beyond the limits of validity of the theory. However, the problem at hand shows up only in the $\alpha<0$ case and thus looks related to the lack of UV completion rather than an issue with effective field theories in general.  Another possibility is that boosts cannot be implemented unitarily if $\alpha<0$.  However, we could not find any particular reason why this should be the case.

Let us look at a rigorous version of the above argument, provided, for example, in the book~\cite{Streater:1989vi}, theorem 4.1.~\footnote{We thank Alessandro Podo for this reference and related discussions.} 
The actual hypothesis in that theorem is that commutators vanish on \emph{open} sets that are mutually spacelike.  In axiomatic field theory, well-defined operators are local fields smeared over spacetime with appropriate spacetime functions of compact support~\cite{Streater:1989vi}. 
It is not sufficient that operators commute on equal time surfaces, say, $[\phi(\vec x, t=0), \phi(0)]=0$. They should commute over a spacetime region of ``finite thickness" enclosing the $t = 0$ plane. This is probably where the standard argument fails.

The actual commutator of a $P(X)$, beyond the first order correction that we managed to compute in~\eqref{eq:commutator1}, is a very complicated operator. Presumably, for $\alpha<0$,  it has support ``almost everywhere" in spacetime.  
The coherent states~\eqref{eq:state} constructed in Sect.~\ref{sec:ExpCOHOsc} seems to support this conjecture. When the continuous label $\mu$ approaches $\mu^2\sim -1/|3\alpha|$,  the sound speed~\eqref{eq:SS} diverges, leading to effectively instantaneous propagation\footnote{One might argue that those states are too close to the cutoff to be seriously considered in the effective theory.  However,  we use them here merely to argue about the formal mathematical structure of the commutator. }. We deduce that the operator-valued commutator is likely non-zero across any open pair of subsets of spacetime. In this respect, it would be analogous to that of a non-relativistic theory, or of a theory with Galilean-boosts symmetry.  While standard equal-time commutation relations can still be imposed at strictly equal time, the commutator becomes non-vanishing at arbitrarily large spatial separations immediately after $t\neq 0$. 
This could be yet another IR diagnosis of a substantial UV obstruction. 

As a word of caution, let us note that the use of regularized distributions from the onset could help confirm the above diagnosis and better understand the connection between the breakdown of microcausality and the hypotheses of the theorem in~\cite{Streater:1989vi}. Any such regulator, however, should preserve both Lorentz invariance and the internal symmetries of the theory, most notably the shift symmetry of the scalar field.  Identifying a suitable regulator consistent with these requirements, and analyzing its implications, is a nontrivial task that we leave to future work.

Finally, one should note that microcausality and Lorentz invariance appear as different ingredients also in the derivation of positivity bounds, see~\cite{Bellazzini:2020cot,EliasMiro:2022xaa, Arkani-Hamed:2020blm} for a complete list of references and~\cite{ Mizera:2023tfe} for a recent review.  While the latter merely restricts amplitudes to depend on Lorentz-invariant kinematic variables $s$, $t$ and $u$, causality further enforces analyticity properties in these variables. As discussed in \cite{Adams:2006sv}, what the above 
$P(X)$ theories with negative signs fail to provide is precisely this analyticity of the amplitude in any would-be UV completion. Thus, although the low-energy amplitude is formally Lorentz invariant, additional input from the UV theory is required to ensure causality.

\section*{Acknowledgments} We warmly thank Brando Bellazzini, Alberto Nicolis, Alessandro Podo,  Andrew Tolley and Filippo Vernizzi for useful conversations and exchanges.  We especially thank Alexander Taskov for his initial collaboration on this project. This work received support from the French government under the France 2030 investment plan, as part of the Initiative d'Excellence d'Aix-Marseille Universit\'e - A*MIDEX (AMX-19-IET-012). This work was supported by the ``action th\'ematique" Cosmology-Galaxies (ATCG) of the CNRS/INSU PN Astro.

\appendix

\section{Commutators with the Yang-Feldman formalism}\label{app_YF}
In this appendix, we show that the interaction picture formalism provides the same perturbative expansion for the interacting field as that obtained by solving the equation of motion for the field operator perturbatively around the free solution.
This second method is known as the Yang-Feldman formalism~\cite{Yang:1950vi}. We show this equivalence in the context of the scalar field theory with quartic contact interactions discussed in the main body of the paper.

We start by writing down the equation of motion of a real scalar field with a quartic interaction
\begin{align}
    (\Box + m^2)\phi &= - \frac{\lambda}{3}\phi^{3}.
    \label{eq:EOMphi4}
\end{align}

By identifying the Green's function of the operator on the left-hand side of Eq~\eqref{eq:EOMphi4} and treating the cubic term as a source, a formal and recursive solution is provided by the following convolution
\begin{equation}
\phi(x) = \phi_0(x) - \frac{\lambda}{3!} \left( \int d^4 z \ G_R(x-z) \phi^3(z) \right).
                \label{eq:YFeq}
            \end{equation}
Eq.~\eqref{eq:YFeq} shows that we have the freedom of choosing the type of Green's function in the convolution. Here, we choose the \textit{retarded} Green's function in order to have a causal evolution. This choice matches the interaction picture formalism.
To evaluate the convolution, we Taylor expand the field
$\phi$ in the coupling constant $\lambda$ according to
\begin{equation}
    \phi = \sum_{n=0}^\infty \lambda^n \phi_n,
\end{equation}
where $\phi_n$ are the coefficients of the series to be determined recursively. In particular, $\phi_1$ iis the first
correction to the free solution where we include the source up to linear order in $\lambda$ corrections, and higher-order terms follow accordingly. 

Then, we plug this relation back into the source term. By keeping in mind that we work with non-commuting fields, we find
\begin{equation}
    \phi^3
    = \phi_0^3 + \lambda \big( \{(\phi_0)^2,\phi_1\} + \phi_0 \phi_1 \phi_0 \big)+ ...
    \label{eq:1YF}
\end{equation}

To obtain the first order correction to the free field, we plug the above expression at the zero order in coupling back into Eq.~\eqref{eq:YFeq}
\begin{align}
    \phi(x) &= \phi_0(x) - \frac{i\lambda}{3!} \int d^4 z\, G_R(x-z) \phi_0^3(z) = \phi_0(x) + \frac{i\lambda}{3!} \int^t_{-\infty} d^4 z\, \Delta(z-x) \phi_0^3(z). 
    \label{eq:phi1}
\end{align}
In the second step, we use the fact that the retarded Green's function is determined in terms of the commutator of the theory according to 
\begin{equation}
    G_R(x-z)=i\theta(t-z^0)[\phi(x),\phi(z)]
\end{equation}
Eq.~\eqref{eq:phi1} coincides with the master formula~\eqref{eq:FirstorderCommutator}
of the interacting picture formalism, expanded at the linear order in $\lambda$, and with $t_*$ chosen to be $-\infty$. 
Note, however, that this last condition is not necessary in general, as Eq.~\eqref{eq:EOMphi4} can be solved with initial conditions chosen so that the field behaves as free at a finite time $t_*$.  Moreover, it is straightforward to prove that the matching between the interaction picture and Yang-Feldman formalism persists also at order $\lambda^2$. This is done by replacing $\phi_1$ entering in Eq.~\eqref{eq:1YF} with the linear order term in $\lambda$ of the solution~\eqref{eq:phi1}, and then plugging everything back into Eq.~\eqref{eq:YFeq}.

In summary, this explicitly shows that the interaction picture formalism provides the interacting field operator as a convolution of operatorial sources and \textit{retarded} Green's functions. 

\section{Coherent states description}
\label{sec:COH}

In this section, we review the properties of the coherent states that we used in the main body of this work. 
Consider the following class of coherent states
\begin{equation}
    |C\rangle=e^{-\ri {f}}|0\rangle,
\end{equation}
with 
\begin{equation}
{f}=\int d^3 {x} \left[\phi^\text{cl}(\vec{{x}}){\pi}(\vec{{x}},0)-\pi^\text{cl}(\vec{{x}}){\phi}(\vec{{x}},0)\right].
\end{equation}
The operator ${f}$ is known as the \textit{shift operator}. 
If we introduce the operator $\mathcal{O}[{\phi}(0,\vec{{x}}),{\pi}(0,\vec{{x}})]$ , which can be any polynomial combination of the field operator and its conjugate momentum $\pi$ or even a transcendental function $\phi$,  the state satisfies the following identity~\cite{Berezhiani:2023uwt}
\begin{equation}
    e^{\ri {f}}\mathcal{O}[{\phi}(\vec{{x}},0),{\pi}(\vec{{x}},0)]e^{-\ri {f}}= \mathcal{O}[{\phi}^\text{cl}(\vec{{x}})+{\phi}(\vec{{x}},0),\pi^\text{cl}(\vec{{x}})+{\pi}(\vec{{x}},0)].
    \label{Shift}
\end{equation}
Therefore, the operator ${f}$ shifts the field operator and the conjugate momentum by the c-number functions $\phi^\text{cl}(\vec{{x}})$ and $\pi^\text{cl}(\vec{{x}})$, hence the name.
Eq.~\eqref{Shift} is independent of the scalar field theory considered and it is derived by only exploiting the canonical commutation relations between the field and its conjugate momentum. By setting $\mathcal{O}={\phi}(\vec{{x}},0)$, the function $\phi^\text{cl}(\vec{{x}})$ is interpreted as the initial value for the one-point function of the field. The same applies to its conjugate momentum ${\pi}$ and the function $\pi^\text{cl}(x)$. Thus, we have
\begin{equation}
    \langle C|\phi(\vec{{x}},0)|C\rangle=\phi^\text{cl}(\vec{{x}}), \qquad  \langle C|\pi(\vec{{x}},0)|C\rangle=\pi^\text{cl}(\vec{{x}})~.
\end{equation}

It is important to note that the choice of the state on which the operator ${f}$ acts, in this case the free Fock vacuum, sets the initial conditions for the connected part of all correlation functions involving more than one power of the fields. Other choices than $|0\rangle$ are possible, such as the interacting vacuum $|\Omega\rangle$ of the full interacting theory, a squeezed vacuum $|S\rangle$, or a non-Gaussian state. The only requirement is that the expectation values of the field operator and its conjugate momentum over the chosen state must vanish.   If this condition is not met, it is possible to redefine ${f}$ and the state on which it acts to enforce this property.\footnote{Notably, a consistent description of the time evolution of the state in the full quantum theory imposes constraints on the form of the state appearing on the right-hand side of Eq.~\eqref{eq:state}, as demonstrated in Ref.~\cite{Berezhiani:2023uwt}.} 
In the main body of the paper, we were interested in expectation values at the leading order in couplings. Within this approximation, the difference between these states becomes negligible.

In Section \ref{sec:ExpCOHOsc}, we consider the expectation value of the operatorial commutator over the coherent state
\begin{equation}
|\mu\rangle=e^{\ri {f}}|0\rangle,\qquad \text{with} \qquad {f}=\mu\left(1+\alpha{\mu^2}\right) \int d^3{x}  \,\phi(\vec{{x}},0)
\label{eq:state}
\end{equation}
We prove that it describes the oscillating background $\phi=\mu t$,  if we neglect loop corrections.
To prove this, we start by evaluating the one-point function of the field operator and the conjugate momentum over $|\mu\rangle,$ at the time of definition of the state $t=0$. 
Using the formula~\eqref{Shift}, these read
\begin{equation}
    \langle \mu|\pi(\vec{{x}},0)|\mu\rangle= \mu\left(1+\alpha {\mu^2}\right), \qquad \langle \mu|\phi(\vec{{x}},0)|\mu\rangle=0, \qquad \langle \mu|\dot{\phi}(\vec{x},0)|\mu\rangle=\mu. 
    \label{eq:ICPi}
    \end{equation}
    The first two relations involving the field operator and the conjugate momentum are exact and do not rely on any loop-expansion. The third identity involving the field velocity  $\dot{\phi}$ 
depends on the relation between the field velocity and the canonical momentum, and only holds if we drop loop corrections.

Equipped with all the initial conditions, we study the time evolution of the one-point functions in the classical limit. To do that, we have to show that the classical expectation value of the conjugate momentum is conserved at all times. Then, using the time-evolved version of Eq.~\eqref{eq:conjugateM}, it would follow that also the expectation value of the classical field velocity is conserved, and 
\begin{equation}
    \langle \mu| \phi(x)|\mu\rangle= \mu t
    \label{eq:1p}
\end{equation}
follows upon time integration.
Let us start by writing down the Heisenberg equation for the conjugate momentum operator, and consider the expectation value over $|\mu\rangle$. This reads
\begin{equation}
 \partial_t \langle \mu| \pi(x)|\mu\rangle=\ri \int {\rm d}^3{z} \langle \mu|[\mathcal{H}(z),\pi(x)]|\mu\rangle\simeq 0
 \label{eq:Heis}
\end{equation}
 The expectation value of the commutator in right-hand side of Eq.~\eqref{eq:Heis} vanishes if we neglect loop corrections because it contains only spacial derivatives of the field.  These operators are not shifted by the shift operator at any time because the state describes a homogeneous configuration.  
In other words, we find that a general contribution to the commutator reads
 \begin{flalign}
   \ri \int {\rm d}^3z \langle \mu|[\mathcal{H}(z,t),\Pi(x,t)]|\mu\rangle&\sim\langle \mu|\partial^i\left\{(\pi^{2m}(x,t) (\partial_k \phi \partial^k \phi)^n \partial_i \phi(x,t)\right\}|\mu\rangle\nonumber \\&\sim \langle \mu|\pi(x,t)|\mu\rangle^{2m-1}\mathcal{O}(\hbar^{n+1}).
\end{flalign}
 with $n\geq 0$ and $m\geq 1$. The $m=0$ contributions coincide with the ones of the vacuum theory and are vanishing due to underlying $Z_2$ invariance of the latter. Having proven the conservation of the classical expectation value of the conjugate momentum, Eq.~\eqref{eq:1p} follows.

\section{$P(X)$ commutator: details of the calculation}
\label{app:EvaluationComm}
In this Appendix, we provide the derivation of the interacting commutator of the $P(X)$ theory, as given by Eq.~\eqref{eq:commutatorDerivative} of Section 3. 
The starting point is  the momentum representation of the interaction picture commutator
\begin{equation}
	\Delta(z_1-z_2)=\int\frac
	{d^3k}{2k (2\pi)^3} \left(e^{-\ri k (z^0_1-z^0_2)}-e^{\ri k (z^0_1-z^0_2)}\right)e^{\ri \vec{k} \cdot (\vec{z}_1-\vec{z}_2)},
\end{equation}
which we use to evaluate Eq.~\eqref{eq:PX1order}.

In particular, we need to evaluate the integrals involving spatial and time derivatives of delta-functions.  We start with the first integral which reads
\begin{flalign}
	\int_0^t d^4z  \partial_i \Delta(z-x) \partial^i \Delta(z)
	&=\int_0^t dz^0\int \frac{\rd^3k}{4(2\pi)^3} \left(e^{-\ri k (2z^0-t)}+e^{\ri k (2z^0-t)}- e^{\ri k t}-e^{-\ri k t}\right)e^{\ri \vec{k}\cdot \vec{x}} \nonumber\\=&-\frac{i}{8\pi^2}\biggr\{\left(\frac{1}{t_-^2-\vec{x}^2}-\frac{1}{t_+^2-\vec{x}^2}\right)-2t^2\left(\frac{1}{ (t^2_--\vec{x}^2)^2 }-\frac{1}{ (t^2_+-\vec{x}^2)^2 }\right)\biggr\},
\end{flalign}
where the imaginary tilt of the time $t_\pm=t(1\pm\ri \epsilon)$ is introduced to have the UV-convergence of the momentum integrals. 
The integral involving time derivatives of delta-functions is evaluated in a similar way and reads 
\begin{flalign}
	\int_{0}^{t} d^4z\partial_0\Delta(z-x)\partial_{0} \Delta(z)=\frac{i}{8\pi^2}\biggr\{\left(\frac{1}{t_-^2-\vec{x}^2}-\frac{1}{t_+^2-\vec{x}^2}\right)+2{t^2}{}\left(\frac{1}{ (t^2_--\vec{x}^2)^2 }-\frac{1}{ (t^2_+-\vec{x}^2)^2 }\right)\biggr\}
\end{flalign}

By plugging the above expressions back into Eq.~\eqref{eq:PX1order}, we obtain Eq.~\eqref{eq:PX1orderfraction}.
The final step is to extract the discontinuities from the above expression to derive Eq.~\eqref{eq:commutatorDerivative}. This is achieved by employing the following operator identities:
\begin{flalign}
&f(x)\delta'(x)=f(0) \delta'(x)-f'(0)\delta(x),
\label{eq:expression1}\\ 
&2 \pi i \delta(x) = \frac{1}{x - i \epsilon} - \frac{1}{x + i \epsilon}
\label{eq:expression3}
\end{flalign}
The first expression is obtained by recalling the identity $f(x)\delta(x)=f(0)\delta(x)$.  Differentiating this identity with respect to 
$x$ then leads directly to Eq.~\eqref{eq:expression1}.  The second expression comes from the well-known application of the Sokhotski–Plemelj theorem on the real-line.
Differentiating the second expression with respect to $x$,  and evaluating the result over $t\pm |\vec{x}|$ provides a third identity that we exploit below
\begin{equation}
\frac{1}{(t_-\pm |\vec{x}|)^2}- \frac{1}{(t_+\pm |\vec{x}|)^2}=\mp 2\ri \pi \frac{\partial}{\partial{|\vec{x}|}} \delta(|\vec{x}|\pm t).
\label{eq:expression2}
\end{equation}

By focussing on the correction proportional to the coupling of Eq.~\eqref{eq:PX1orderfraction}, we write
\begin{flalign}
        &\left(\frac{1}{t_-^2-\vec{x}^2}-\frac{1}{t_+^2-\vec{x}^2}\right)\nonumber-t^2\left(\frac{1}{(t_-^2-\vec{x}^2)^2}-\frac{1}{(t_+^2-\vec{x}^2)^2}\right)\\
        &\quad {	= \frac{\pi i}{|\vec{x}|}\left(\delta(|\vec{x}|-t) -\delta(|\vec{x}|+t)\right)-  \frac{t}{4 |\vec{x}|}\left(\frac{1}{(t_- -|\vec{x}|)^2}-\frac{1}{(t_+ -|\vec{x}|)^2}-\frac{1}{(t_- +|\vec{x}|)^2}+\frac{1}{( t_ + + |\vec{x}|)^2} \right) } \nonumber \\
        &\quad {= \frac{\pi i}{|\vec{x}|} \left\{   \left(\delta(|\vec{x}|-t)-\delta(|\vec{x}|+t)\right)- \frac{t}{2} \frac{\partial}{\partial |\vec{x}|}\left(\delta(|\vec{x}|-t)+\delta(|\vec{x}|+t)\right) \right\} },
\end{flalign}
where in the first  and second step we used Eq.~\eqref{eq:expression3} and  Eq.~\eqref{eq:expression2}. Finally, by introducing back overall coefficients and using Eq.~\eqref{eq:expression1}, we find
\begin{flalign}
	=&\frac{\alpha\mu^2}{2i\pi^2|\vec{x}
    |} \biggr\{-\big(\delta(|\vec{x}|-t)-\delta(|\vec{x}|+t)\big)+\frac{t}{2}\frac{\partial}{\partial |\vec{x}|}\big(\delta(|\vec{x}|-t)+\delta(|\vec{x}|+t)\big)\biggr\}\nonumber\\=&\frac{\alpha\mu^2}{4i\pi^2|\vec{x}|} \left\{-\big(\delta(|\vec{x}|-t)-\delta(|\vec{x}|+t)\big)+|\vec{x}|\frac{\partial}{\partial |\vec{x}|}\big(\delta(|\vec{x}|-t)-\delta(|\vec{x}|+t)\big)\right\}\nonumber\\
    =&\frac{\alpha\mu^2}{2i\pi^2}\text{sign}(t) |\vec{x}|\frac{\partial}{\partial |\vec{x}|} \delta(\vec{x}^2-t^2)
\end{flalign}
which can be reduced to the first line of Eq.~\eqref{eq:commutatorDerivative}.

\bibliographystyle{unsrt}
\bibliography{refs.bib}

\end{document}